\documentclass[twocolumn]{aastex62}
\usepackage{lineno}
\usepackage{graphics,epsf}
\usepackage{amsmath}                
\usepackage{amsfonts}               
\usepackage{amssymb}                
\usepackage{epsfig}                 
\usepackage{appendix}
\usepackage{graphicx}
\usepackage{float}
\usepackage{color}
\usepackage{multirow}
\usepackage{colortbl}
\usepackage[para,online,flushleft]{threeparttable}

\newcommand{\kms}{{~\rm km\; s^{-1}}}
\newcommand{\cc}{{~\rm cm^{-3}}}

\newcommand{\cm}{{~\rm cm}}

\newcommand{\km}{{~\rm km}}
\newcommand{\s}{{~\rm s}}

\newcommand{\g}{{~\rm g}}

\newcommand{\K}{{~\rm K}}
\newcommand{\erg}{{~\rm erg}}

\begin{document}

\title{Post-explosion positive jet-feedback activity in inner ejecta of core collapse supernovae}


\author{Muhammad Akashi}
\affiliation{Department of Physics, Technion, Haifa, 3200003, Israel; akashi@physics.technion.ac.il; soker@physics.technion.ac.il}
\affiliation{Kinneret College on the Sea of Galilee, Samakh 15132, Israel}

\author{Noam Soker}
\affiliation{Department of Physics, Technion, Haifa, 3200003, Israel; akashi@physics.technion.ac.il; soker@physics.technion.ac.il}

\begin{abstract}
We conduct three-dimensional hydrodynamical simulations of weak jets that we launch into a core collapse supernovae (CCSNe) ejecta half an hour after the explosion and find that the interaction of the fast jets with the CCSN ejecta creates high pressure zones that induce a backflow that results in mass accretion onto the newly born neutron star.
In cases of weak jets, a total power of $\approx 10^{45}-10^{46} \erg$,
the backflow mass accretion might power up to an order of magnitude more energetic jets. In total, the jets of the two post-explosion jet-launching episodes have enough energy to influence the morphology of the very inner ejecta, a mass of $\approx 0.1 M_\odot$.    
Our results imply that in some, probably a minority of, CCSN remnants the very inner regions might display a bipolar structure that results from post-explosion weak jets. The regions outside this part might display the morphology of jittering jets. 
\end{abstract}


\section{Introduction} 
\label{sec:intro}

In core-collapse supernovae (CCSNe) that form a neutron star (NS) the inner regions of the core of the progenitor star collapse in about a second to form the NS (e.g., \citealt{Hegeretal2003}). This collapse releases a gravitational energy of $>{\rm few} \times 10^{53} \erg$ (e.g., \citealt{Janka2012}). Neutrinos carry most of this energy (e.g., \citealt{Fujibayashietal2021, Nakamuraetal2022} for a recent studies). The exploding outer core and envelope (if exists) carry the rest of the energy in the forms of thermal energy and kinetic energy. The very inner parts of the collapsing core form a dense proto-NS as they come to rest at nuclear densities in the center and a shock wave propagates out (e.g., \citealt{BurrowsVartanyan2021}). This shock wave stalled at about $100 \km$ from the center (e.g., \citealt{Bruennetal2013}), and the rest of the collapsing matter passes through the stalled shock wave and settled onto the newly born NS (e.g., \citealt{Janka2012}). 
   
In recent years studies discuss two theoretical explosion mechanisms that are based on  gravitational energy. In the delayed neutrino mechanism neutrinos heat the post-shock collapsing matter and revive the stalled shock  (e.g.,  \citealt{BetheWilson1985, Hegeretal2003, Nordhausetal2012, CouchOtt2013, Bruennetal2016, Jankaetal2016R, Mulleretal2019Jittering, BurrowsVartanyan2021, Fujibayashietal2021, Bocciolietal2022, Nakamuraetal2022}).
In the jittering jets explosion mechanism the newly born NS (or black hole; BH) launches jets that deliver the energy to the ejecta (e.g., \citealt{Soker2010, PapishSoker2011, GilkisSoker2014, PapishSoker2014Planar, Quataertetal2019, Soker2020RAA, ShishkinSoker2021, AntoniQuataert2022, ShishkinSoker2022, Soker2022}).

{{{ The main advantages of the delayed neutrino mechanism are that neutrino heating must take place, it is very important even in the jittering jets explosion mechanism (e.g.,  \citealt{Soker2022Boosting}), and that numerical simulations achieve explosions in some cases (e.g. \citealt{Bolligetal2021, BurrowsVartanyan2021}). There are some disadvantages of the delayed neutrino mechanism (e.g., \citealt{Kushnir2015, Papishetal2015}), the main two being that in some cases explosion does not occur (e.g., \citealt{Burrowsetal2020}), and that the maximum explosion energy this mechanism can give is $\simeq 2 \times 10^{51} \erg$ (e.g. \citealt{Fryeretal2012, Sukhboldetal2016}). The main advantage of the jittering jets explosion mechanism is that it accounts for the imprints of jets in CCSN remnants (e.g., \citealt{GrichenerSoker2017}) and that it can account for any explosion energy (e.g., \citealt{Gilkisetal2016Super}). Its main disadvantage is that there are yet no numerical simulations that show its feasibility. }}}

The jittering jets explosion mechanism {{{ is based on a }}}  negative feedback mechanism in the sense that the jets remove mass that serves as the reservoir for the accretion disk than launches the jets (see review by \citealt{Soker2016Rev}). The negative jet feedback mechanism, {{{ which is a critical component of the jittering jets explosion mechanism, }}} explains why the explosion energy of most CCSNe is of the order of the binding energy of the star (mainly contribution from the binding energy of the pre-collapse core) to several times that energy. High efficiency, {{{ i.e., a high fraction of the initial jets' energy is channelled to unbind the exploding parts of the star, }}} occurs when the jets jitter or precess, i.e., change their directions on short timescales of $\approx 0.1 
\s - {\rm few}\times 0.1 \s$, because in those cases the jets interact with core material in all directions. 
  
When the jets maintain a fixed axis rather than jittering axes the situation is substantially different (\citealt{Gilkisetal2016Super, Soker2017RAA}). In the fixed axis explosion process the explosion is still driven by jets, but there is only a little jittering. Technically it is part of the jittering jets explosion mechanism, but jittering is very small and occurs around a fixed axis. {{{ Although the pre-collapse core has a large amount of angular momentum, the pre-collapse convection and instabilities above the NS lead to angular momentum fluctuations on top of the fixed pre-collapse angular momentum of the core. }}} In the fixed axis explosion process the jets interact only with core and envelope gas along the polar directions. Matter from around the equatorial plane flows in and increase the mass of the newly born NS. The final mass of the central compact object is a substantial fraction of the pre-collapse core mass, and so the NS collapses to form a BH. The fixed axis explosion takes place when the pre-collapse core is rapidly rotating, most likely by a merger with a star in a common envelope evolution. 

There are many theoretical and observational studies of jet-driven CCSNe (e.g.,  magnetorotational mechanism) that take the jets' axis to be fixed (e.g., \citealt{Khokhlovetal1999, Aloyetal2000, Maedaetal2012, LopezCamaraetal2013, BrombergTchekhovskoy2016,  Nishimuraetal2017, Perleyetal2021}). The jittering jets explosion mechanism is different in that (1) it assumes that most (or even all) CCSNe are driven by jets, not only those that result from rapidly-rotating progenitors, (2) it is based on a negative feedback mechanism (see review by \citealt{Soker2016Rev}), and (3) even in the case of a large angular momentum of the pre-collapse core the mechanism allows for some jittering due to stochastic angular momentum accretion.
{{{ The source of the stochastic angular momentum of the accreted gas are instabilities in the post-stalled-shock zone, mainly the spiral standing accretion shock (SASI) instability (e.g.,  \citealt{BlondinMezzacappa2007, Kazeronietal2017}) seeded by large perturbations from the pre-collapse inner convective zones of the core (e.g., \citealt{ShishkinSoker2022}). }}}

In the jittering jets explosion mechanism there are no failed CCSNe (e.g., \citealt{Gilkisetal2016Super, Soker2017RAA, AntoniQuataert2022}), a claim that received a strong support with the new observational finding by \cite{ByrneFraser2022} {{{ that failed CCSNe are very rare, or even do not exist. }}} Even if a BH forms at the center it launches very energetic jets. In other words, while according to the delayed neutrino mechanism the formation of a BH occurs in a failed CCSN that yields only a faint transient event (e.g., \citealt{Nadezhin1980, LovegroveWoosley2013}), according to the fixed axis explosion process of the jittering jets explosion mechanism the formation process of a BH in a CCSN is accompanied by the most energetic CCSN explosions (e.g., \citealt{Gilkisetal2016Super, Soker2017RAA}). For example, according to the picture of the jittering jets explosion mechanism the transient event that some attribute to a failed CCSN (e.g., \citealt{Adamsetal2017, Basingeretal2021}) might be a supernova impostor, like a type II intermediate luminosity optical transient \citep{KashiSoker2017, Soker2021IILOT, Bearetal2022}. 
 
In the present study we explore some properties of the flow that jets induce in the inner ejecta of a CCSN about half an hour after explosion.  
In section \ref{sec:Numerics} we describe the numerical setting, and in section \ref{sec:Results} we present the results. We summarize this study in section \ref{sec:Summary}. 

\section{Numerical procedure} 
\label{sec:Numerics}

\subsection{Three-dimensional hydrodynamical procedure}
\label{subsec:3Dhydro}

We use version 4.6.2 of the adaptive-mesh refinement (AMR) hydrodynamical FLASH code \citep{Fryxell2000} in three dimension (3D). We turn off the radiative cooling since the regions we simulate are optically thick. 
We include radiation pressure, electrons pressure, and ions pressure as we assume an adiabatic index of $\gamma=5/3$. 
We set outflow conditions on all boundary surfaces of the 3D grid. We use resolution with 7 refinement levels, i.e., the ratio of the length of the sides of the grid to the shortest cell in the grid is $2^{9}$. {{{ The finest cell in the grid has a side of $0.0234 R_{\odot}$. In addition, for one case we compare results of simulations with 6 refinement levels and with 8 refinement levels. As we describe in section \ref{subsec:BackFlow} we reach resolution convergence with 7 refinement levels. }}} 


We simulate the whole space with a total size of the Cartesian  numerical grid of $(12 R_\odot)^3$, i.e., $(L_x,L_y,L_z) = \pm 6 R_\odot$.
{{{ We simulate the entire space, rather than only one half of the symmetry plane $z=0$, to avoid numerical artifacts at the symmetry plane had we simulated only one half of the interaction region. We will present the flow structure in the  entire computational grid to emphasize the bipolar structure that the jets shape and the flow in the equatorial plane. }}}


We take for the mass of the CCSN ejecta $M_{\rm ej} = 5 M_{\odot}$ and for its kinetic energy $E_{\rm SN} = 4 \times 10^{51} \erg$. {{{ These values correspond to an energetic stripped-envelope CCSN, which is compatible with the presence of a fall back material that feeds jets after explosion. }}} We take the initial density and velocity profiles of the ejecta from \cite{SuzukiMaeda2019} (their equations 1-6, with $l=1$ and $m=10$). Namely, we assume that before the launching of jets the ejecta outflows in its terminal velocity $\hat r r/t_{\rm e}$, where $t_{\rm e}$ is the time from explosion. {{{ However, for numerical reasons (see below) we add a constant magnitude to this velocity such that in the numerical grid the initial ejecta velocity is $\vec v_{\rm ejecta}(t)=[\Delta v_{\rm num} + r/t_{\rm e} ] \hat r= [2500 \km \s^{-1} + (r/1800 \s) ]\hat r$. }}} 
The density profile is 
\begin{equation}
\rho (r, t) = \begin{cases}
        \rho_0 \left( \frac{r}{t_{\rm e} v_{\rm br}} \right)^{-1} 
        & r\leq t v_{\rm br}
        \\
        \rho_0 \left( \frac{r}{t_{\rm e} v_{\rm br}} \right)^{-10} 
        & r>t v_{\rm br}, 
        \end{cases}
\label{eq:density_profile}
\end{equation}
    \newline
where   
\begin{eqnarray}
\begin{aligned} 
& v_{\rm br} = \left( \frac{20}{7} \right)^{1/2} \left( \frac {E_{\rm SN}}{M_{\rm ej}} \right)^{1/2} 
=1.07 \times 10^ 4 
\\& 
\times \left( \frac {E_{\rm SN}}{4 \times 10^{51} \erg} \right)^{1/2}
\left( \frac {M_{\rm ej}}{5 M_\odot} \right)^{-1/2}
\km \s^{-1} ,
\end{aligned}
\label{eq:vbr}
\end{eqnarray}
{{{  is the velocity at the break that separates the inner and outer parts of the ejecta, }}} and
\begin{equation}
\rho_0 = \frac {7 M_{\rm ej}}{18 \pi v^3_{\rm br} t^3_{\rm e}} = 1.71\times 10^{-4} \g \cc .
\label{eq:rho0}
\end{equation}

Our grid includes only the inner part of ejecta, i.e., $v \ll v_{\rm br}$. The initial temperature of the ejecta in our simulations is $T_{\rm 0,ej} = 10^4 \K$, and so the initial Mach number of the ejecta {{{ is Mach$> \Delta v_{\rm num}/C_{\rm s} = 150$, where  $\Delta v_{\rm num} = 2500 \km \s^{-1}$ is the extra numerical velocity (see below) and $C_{\rm s}$ is the sound speed.  }}}

\begin{figure}
\includegraphics[trim= 3.5cm 8.5cm 0.0cm 6.5cm ,clip, scale=0.55]{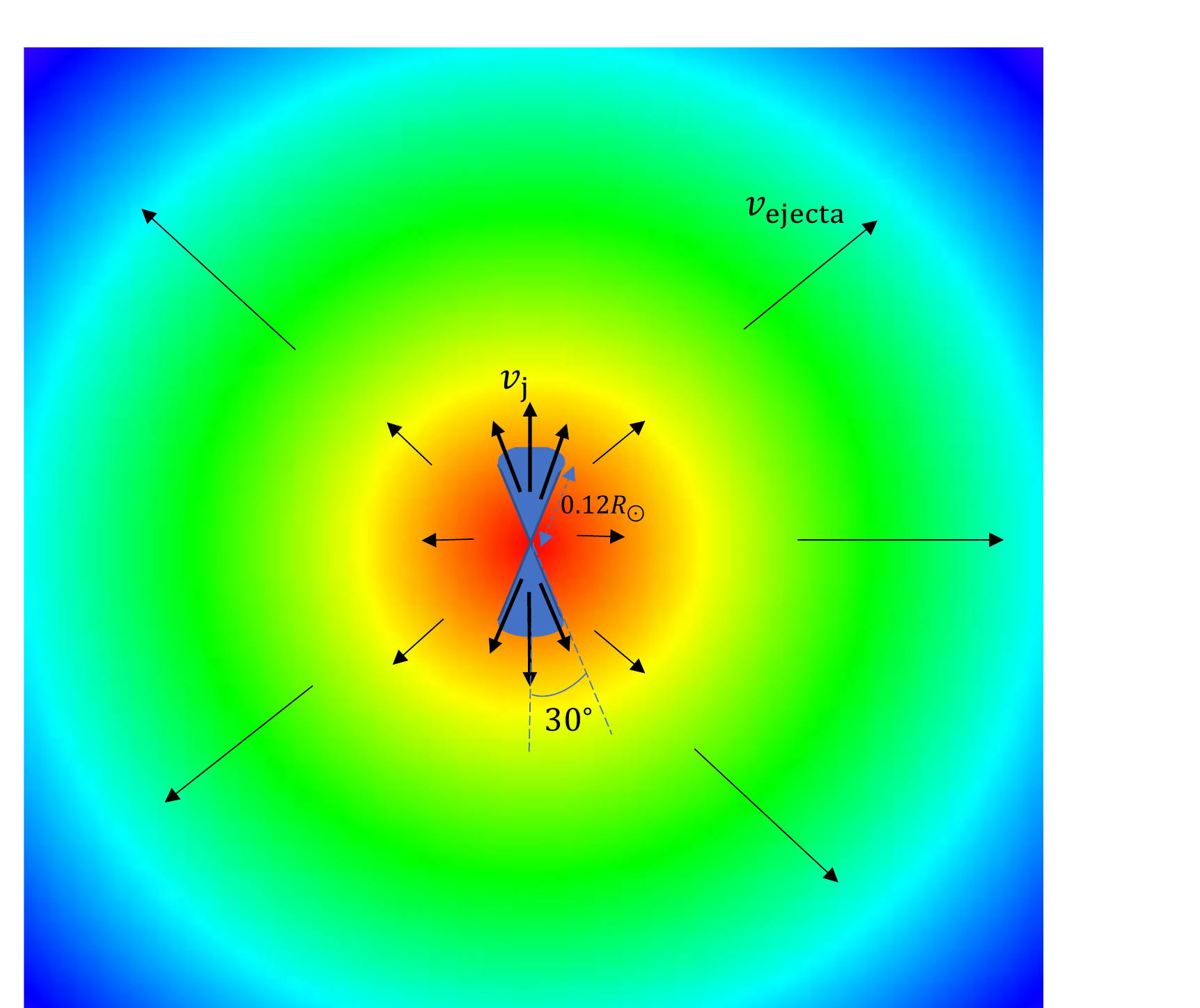} \\
\caption{ {{{ Schematic drawing of the initial flow in the meridional plane $y=0$. We insert the jets with a velocity of $\vec v_{\rm j} = (5 \times 10^4 \km \s^{-1}) \hat r$ inside the two opposite cones. The $z$-axis is along the symmetry axis of the jets. The jets are active for either $\Delta t_{\rm j} =10 \s$ or for $\Delta t_{\rm j} =100 \s$. The initial velocity of the ejecta is $\vec v_{\rm ejecta}(t)=[2500 \km \s^{-1} + (r/1800 \s) ]\hat r$ (see text for the numerical additional velocity of $\Delta v_{\rm num}=2500 \km \s^{-1}$).  }}}
}
	\label{fig:schematic}
\end{figure}

We inject the jets into the grid inside two opposite cones with a length of $0.12 R_\odot$ and a half-opening angle of $\alpha = 30^{\circ}$, {{{ as we present in Fig. \ref{fig:schematic}.  At each time step during the jet-launching episode, which lasts for 10 seconds or 100 seconds, we set the velocity inside the entire volume of the two cones to be radial with a constant magnitude $\vec v_{\rm j} = v_{\rm j} \hat r$, }}} and take for the density of the jets inside the cones  $\rho_j (r) = \dot{M}_{2j} / [4\pi r^{2} v_j (1-\cos \alpha)]$.

At the beginning of the simulations, $t=0$, the two opposite cones are filled with the jets material, {{{ as in Fig. \ref{fig:schematic}. }}} The jets are launched half an hour after the CCSN explosion, as in many CCSNe the fall back material might form an accretion disk about half an our after the explosion (e.g., \citealt{Ronchietal2022}). In our timeline the explosion occurs at  $t_{\rm exp}=-0.5 h$. The jet-launching period lasts for $\Delta t_{\rm j} =10 \s$ in most simulations, beside one simulation where we take $\Delta t_{\rm j} =100 \s$. {{{ We take these timescales for being much longer than the expected jet-activity period during explosion, about one second, but much shorter than the time after explosion, 1800 second. }}} The initial jets' velocity magnitude is $v_{\rm j}=5 \times 10^4 \kms$ {{{ and the mach number is 3300. We expect the jets from a NS to be about 2-3 times faster, but because of numerical limitations we take a somewhat lower velocity. }}} We vary the mass outflow rate in the jets, as we list for the two jets combined, $M_{\rm 2j}$, in the second column of Table \ref{Table:cases}. {{{ In the third column we list the initial energy (basically kinetic) energy of the jets $E_{\rm 2j}=0.5 M_{\rm 2j} v^2_{\rm j}$. We simulate these cases as our goal is to study weak post-explosion jets. The masses we take in the jets with their velocity span an energy range from six to two orders of magnitude below the explosion energy.  }}}
\begin{table}
\tiny
\centering
\begin{tabular}{|c|c|c|c|c|c|c|}
\hline
Sim. & $M_{\rm 2j}$ & $E_{\rm 2j}$ & $\Delta t_{\rm j}$ & $M_{\rm acc}$ & $E_{\rm 2j, next}$ \\ 
           & $M_\odot$ & $\erg$ & $\sec$ & $M_\odot$ & $\erg$\\ 
 \hline 
SN1       & $1.6 \times 10^{-7}$  & $4 \times 10^{45}$ & $10$ & $8.35 \times 10^{-7}$ & $2.6 \times 10^{46}$ \\ \hline
SN2      & $1.6 \times 10^{-6}$ & $4 \times 10^{46}$ & $10$ & $2.6 \times 10^{-6}$ & $8.0 \times 10^{46}$\\ \hline
SN3       & $1.6 \times 10^{-5}$  & $4 \times 10^{47}$  & $10$ & $6.3 \times 10^{-6}$ & $1.9 \times 10^{47}$\\ \hline
SN4      & $1.6 \times 10^{-5}$  &$4 \times 10^{47}$  & $100$ & $4.6 \times 10^{-6}$ & $1.4 \times 10^{47}$\\ \hline
SN5  & $1.6 \times 10^{-4}$ & $4 \times 10^{48}$ & $10$ & $10^{-5}$ & $3.1 \times 10^{47}$\\ \hline
SN6 & $1.6 \times 10^{-3}$ & $4 \times 10^{49}$ & $10$ & $9.6\times 10^{-6}$ & $3.0 \times 10^{47}$ \\ \hline

\end{tabular}
\caption{Summary of the 6 simulations (Sim.) we present in the paper. {{{ In all simulations the CCSN ejecta mass and explosion energy are $M_{\rm ej}=5 M_\odot$ and $E_{\rm SN} = 4\times 10^{51} \erg$, respectively. }}} The columns list, from left to right and for each simulation, its name, the total mass of the two jets, the total energy of the two jets $E_{\rm 2j} = 0.5M_{\rm 2j} v_{\rm j}^2$, the duration of the jet-launching episode, the accreted mass onto the inert core until $t=1000 \s$, and the kinetic energy of the jets in the next jet-launching episode (see section \ref{subsec:BackFlow}).}
\label{Table:cases}
\end{table}

We do not include the self-gravity of the ejecta, but we do include the gravity of the newly born NS as point mass of $M_{\rm NS}=1.4 M_\odot$ at the center. In a test simulation without jets we find that, because of the central gravity, the gas surrounding the centre falls back even without launching jets. As our goal is to find the effect of the jets on the backflow accretion, we add a velocity of $\Delta v_{\rm num} = 2500 \kms$ to the initial velocity profile based on \cite{SuzukiMaeda2019}.
Namely, we take for the initial velocity of the ejecta in the  simulations The initial velocity of the ejecta is $\vec v_{\rm ejecta}(t)=[2500 \km \s^{-1} + (r/1800 \s) ]\hat r$. 
Because of this additional velocity to overcome numerical problems, we might underestimate the fallback accretion due to the effects of the jets.  

{{{ Overall, the independent physical parameters and variables of the simulations are the explosion energy and the ejecta mass (we take a spherical CCSN explosion), the starting time and ending time of the jet activity relative to the explosion time, the mass and velocity of the (highly supersonic) jets at outlet, and the half opening angle of the jets. Because the jets and ejecta have very high Mach numbers the initial temperature, $10^4 \K$, does not play a noticeable role in this study. The numerical variables are the resolution of the numerical grid, the length of the cones into which we inject the jets, and the extra ejecta velocity $\Delta v_{\rm num}$. }}}   

\subsection{The accretion procedure}
\label{subsec:AccretionProcedure}
We fixed an inert core at the centre of the grid in which we set a very low pressure and zero velocity. The radius of the inert core is $R_{\rm inert} = 0.1 R_\odot$.
We calculate the accreted mass at each time step by examining the density and velocity inside a thin spherical shell of $0.12R_\odot<r<0.14R_\odot$. 
At each time step we sum over the quantity $(\rho_{\rm cell}-\rho_0)V_{\rm cell}$ in each cell with an inflow, i.e., $\vec{v} \cdot \vec{r} < 0$, where $\rho_{\rm cell}$ is the density in the cell, $V_{\rm cell}$ is the volume of the cell, and $\vec{v}$ is the velocity of the gas in the shell. $\rho_0$ is a very low density that prevents numerical problems had we set $\rho_0=0$.  
At the end of each time step we reset the density in the cells of the thin shell to be $\rho_0$ and the velocity of these cells to be zero.

\section{Results} 
\label{sec:Results}
\subsection{The flow structure} 
\label{subsec:Flow}
 
\begin{figure*}
\includegraphics[trim=-1.cm 2.cm 0.0cm 0.0cm ,clip, scale=0.80]{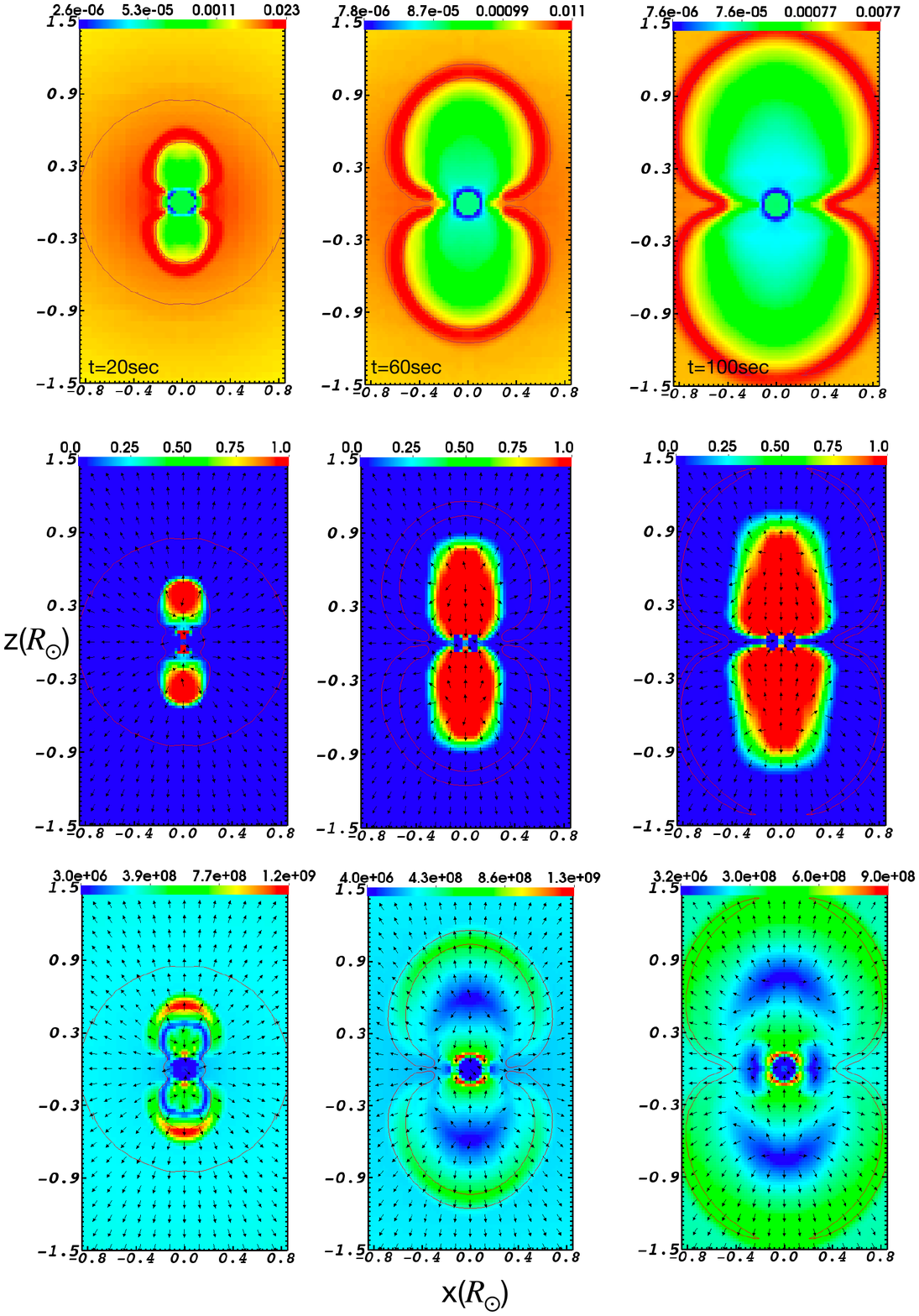} \\
\caption{Density (upper row; color-bar in $\g \cm^{-3}$), jet's tracer (middle row; $0 \le \xi \le 1$ by color-bar), and velocity (bottom row; color-bar in $\cm \s^{-1}$) maps at three times of $20 \s$, $60 \s$, and $100 \s$, from left to right, respectively, of simulation SN3. {{{ The density contour in all panels is for $\rho = 0.005 \g \cm^{-3}$, to emphasise the high-density boundary of the two bubbles. }}} Note that the jets activity ended at $t=10 \s$. }
	\label{fig:six_maps}
\end{figure*}

We set $t=0$ when we start to launch the jets, which is half an hour after explosion (section \ref{subsec:3Dhydro}). The two opposite jets that we launch with an initial velocity of  $v_{\rm j}=5 \times 10^4 \kms$ expand into the slower ejecta and shock it. In Fig. \ref{fig:six_maps} we present the flow structure at three times after the jet-launching episode has ended, from left to right, of simulation SN3 (see table). In that figure we present the density, jets' tracer, and velocity maps in the meridional plane $y=0$. The jets' tracer is an artificial Lagrangian variable that flows with the jets' material and represents the fraction of jets' material by mass in each grid cell. Namely, its value is $\xi=1$ for a pure jet material and $\xi=0$ if there is no mass in the cell that originated in the jets. In other cells its value is $0 < \xi < 1$.  

\begin{figure*}
\includegraphics[trim=0.0cm 9.0cm 0.0cm 1.0cm ,clip, scale=0.8]{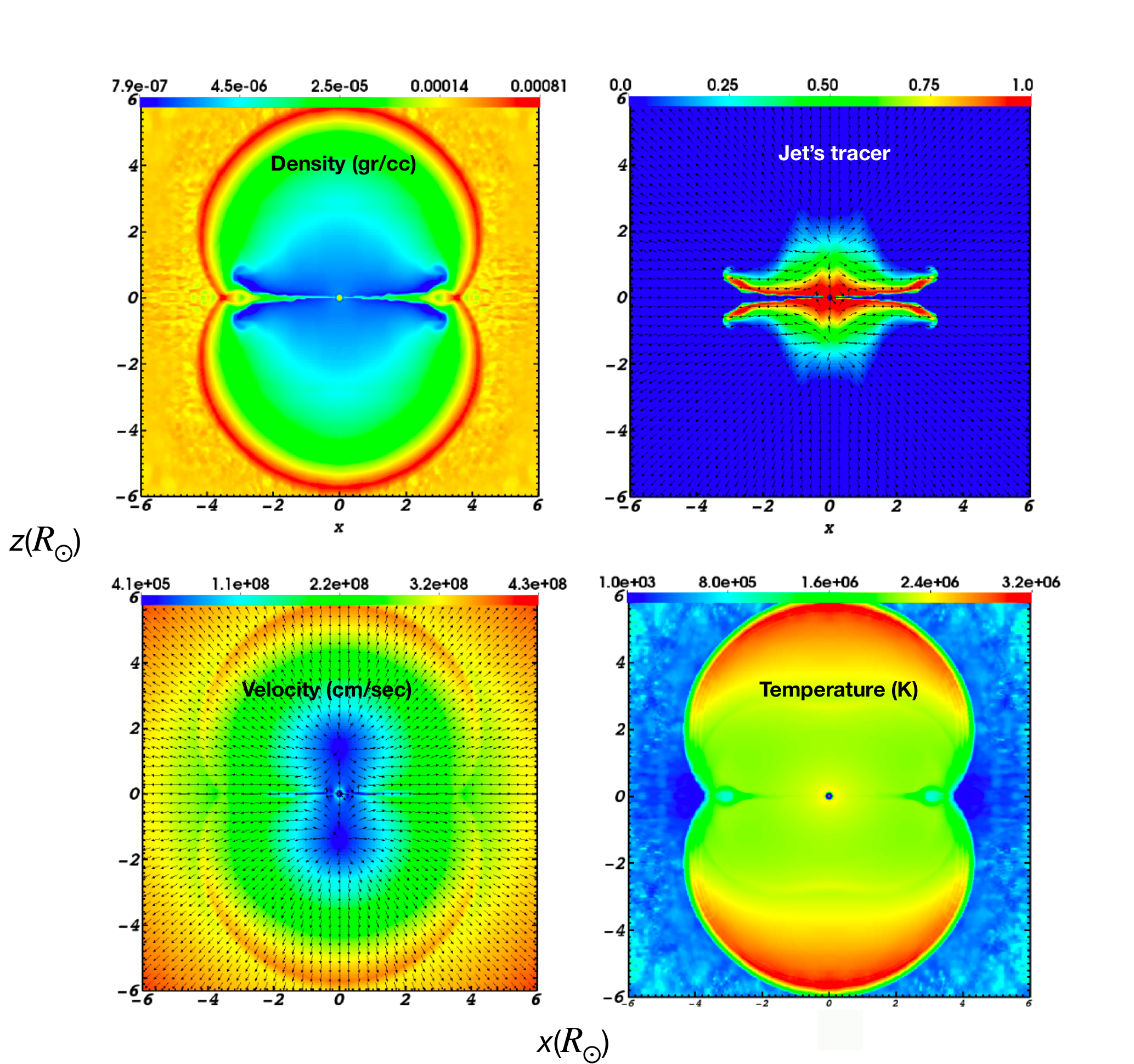} \\
\caption{Density (upper-left; color-bar in $\g \cm^{-3}$), jet's tracer (upper-right; $0 \le \xi \le 1$ by color-bar), velocity (bottom-left; color-bar in $\cm \s^{-1}$), and temperature (bottom-right; color-bar in K) maps at $t=800 \sec$ for simulation SN3. }
	\label{fig:four_maps}
\end{figure*}

The jets-ejecta interaction forms two opposite hot low-density bubbles (green regions in the upper row of Fig. \ref{fig:six_maps}), where the inner regions of the bubbles along and near the symmetry axis are filled with the post-shock jets material (red and green regions in the middle row of Fig. \ref{fig:six_maps}) and the outer regions of the bubbles are filled with the post-shock ejecta. The shocked materials of the jets and ejecta form the `cocoon' around the jets. After the jets ceased, the cocoon fills the entire bubble. In Fig. \ref{fig:four_maps} we present the density, tracer, velocity, and temperature maps in the meridional plane of simulation SN3 at $t=800 \s$. Note that the scale of the meridional plane in Fig. \ref{fig:four_maps} is much larger than that in Fig. \ref{fig:six_maps}. 

{{{ From Fig. \ref{fig:four_maps} we learn that a long time after the jet activity has ceased the morphology has changed. The density map (upper-left panel of Fig. \ref{fig:four_maps}) shows that the two bubbles have merged to form one large (almost) prolate bubble. The tracer of the jets (upper right) shows that the flow has pushed a large fraction of the shocked jets' material towards the symmetry plane. Indeed, the velocity map has changed due to a large backflow (lower left panel), that leads to accretion and pushes the jets' material towards the symmetry plane. The symmetry axis of the jets still has its imprints, e.g., in the elongated one bubble and the two opposite high-temperature caps (lower right panel).
}}}

The region around the head of a jet during the jet-activity phase is of high pressure and it accelerates material out from that region. In particular, there is a large region where the shocked jets' material flows backwards towards the center. A fraction of this back-flowing gas is accreted by the newly born NS at the center. Finding the backflow accretion rate is the main goal of this study. 

\subsection{The backflow accretion} 
\label{subsec:BackFlow}

\begin{figure*}
\includegraphics[trim=0.1cm 5.5cm 0.cm 6.0cm ,clip, scale=0.8]{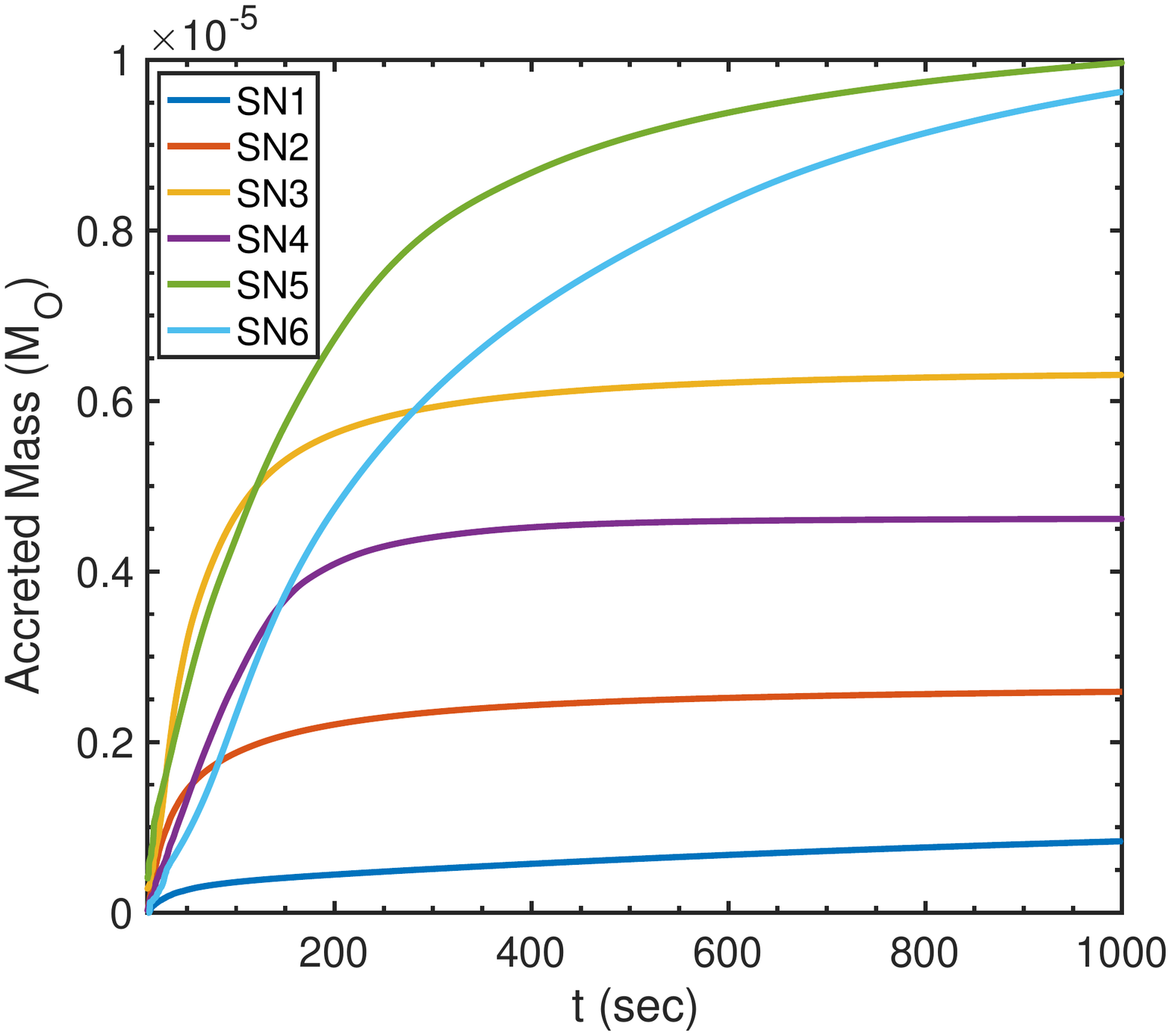} \\
\caption{Backflow accreted mass (in $M_\odot$) onto the inert core as a function of time for the six simulations we perform here (Table \ref{Table:cases}). {{{ In the inset we list the the names of the different simulations.}}} }
	\label{fig:temp}
\end{figure*}

We calculate the accreted mass as a result of the backflow as we described in section \ref{subsec:AccretionProcedure}. In Fig. \ref{fig:temp} we present the accreted mass as function of time until $t=1000 \s$, and list the total accreted mass by that time in the fifth column of Table \ref{Table:cases}. 
{{{ Although the initial mass in the jets span four orders of magnitude, the accreted mass spans about one order of magnitude, from  $M_{\rm acc} = 8.35\times 10^{-7}M_{\odot}$ to $M_{\rm acc} = 10^{-5} M_{\odot}$. }}}

From simulation SN1 (the weakest jets) to SN5 the amount of backflow accreted mass increases with increasing jets' energy. However, {{{ the amount of accreted mass increases by about an order of magnitude while the energy increases by three orders of magnitude from simulation SN1 to simulation SN5. }}} The amount of accreted mass decreases then in the simulation with the most energetic jets (simulation SN6; pale blue in Fig. \ref{fig:temp}). {{{ Our qualitative explanation for this decrease is as follows. The collision of a jet with the ejecta channels kinetic energy to thermal energy, forming a high-pressure zone. This high pressure zone accelerates material backward towards the center. The interaction becomes stronger as the jets are more energetic, explaining the increase in accreted mass from simulation SN1 to simulation SN5. However, with the higher energy in our simulations comes larger momentum in a jet. The momentum is conserved, and therefore higher momentum acts against a backflow. This explains the relatively moderate increase in mass accreted with the increase in the energy of the jets. In going from simulation SN5 to simulation SN6 the change in the effect of momentum against backflow is larger than the change in the effect of jets' energy for backflow, and so the amount of accreted mass decreases somewhat.  }}} 
 
We also note the not-too-large differences between simulation SN3 and simulation SN4, both of which have the same total energy and mass in the jets, but in simulation SN4 the jet-launching episode lasts for $\Delta t_{\rm j}=100 \s$ instead of $\Delta t_{\rm j} =10 \s$.

{{{ For the parameters of simulation SN3 we conducted two extra simulations to check resolution convergence. One extra simulation had a low resolution of only 6 refinement levels, rather than 7 refinement levels as in all the simulations that we present in Table \ref{Table:cases}, and one simulation had a higher resolution with 8 refinement levels. We find the amounts of accreted mass for these three cases to be $M_{\rm acc,6}=7.8 \times 10^{-6} M_\odot$, $M_{\rm acc}=6.3 \times 10^{-6} M_\odot$ (as in the fourth row/fifth column of Table \ref{Table:cases}), and $M_{\rm acc,8}=6.2 \times 10^{-6} M_\odot$, respectively. We compared also the flow maps of the two higher resolution simulations and could not notice significant differences.
We conclude that we reached a resolution convergence to about 2 per cents with 7 refinement levels. }}}

To assess the importance of this backflow we calculate the plausible energy of the jets that this backflow might launch. We assume that the accreted mass has sufficient amount of angular momentum to form an accretion disk and launch jets. {{{ This is a reasonable assumption because we assume that this CCSN is more energetic than typical CCSNe, and that the reason for that is the rapid rotation of the pre-collapse core. This in turn implies that the ejecta has sufficient specific angular momentum to form an accretion disk.  }}} We also assume that these jets carry about 10\% of the gravitational energy that the accretion process releases. The energy in the jets of the next jet-launching episode might be therefore 
\begin{equation}
E_{\rm 2j,next} = 0.1 \frac{G M_{\rm NS}M_{\rm acc}}{R_{\rm NS}} ,
\label{eq:Enext}
\end{equation}
where $M_{\rm NS}=1.4 M_\odot$ and $R_{\rm NS} = 12 \km$ are the mass and radius of the NS, respectively.  
We list this energy for each of the six simulations in the last column of Table \ref{Table:cases}. 

From the last column of Table \ref{Table:cases} we learn that even weak jets might induce the next jet-launching episode to be as powerful as the original jets are or even more. {{{ Because the second jet-launching episode starts within $\approx 100 \s$, the two new jets expand into the two low-density bubbles  that the first jets shaped (upper-right panel of Fig. \ref{fig:six_maps}). The jets of the second jet-launching episode will not lead to a new bipolar structure, but rather the effect of these jets would be to act as if the first jet-launching episode is more energetic. We make this claim under the reasonable assumption that the two launching episodes share the same angular momentum (symmetry) axis.   }}} 
   
\section{Discussion and Summary} 
\label{sec:Summary}

We injected weak jets, {{{ from $E_{\rm 2j} = 4 \times 10^{45} \erg = 10^{-6} E_{\rm SN}$ to $E_{\rm 2j} = 4 \times 10^{49} \erg = 0.01 E_{\rm SN}$ (Table \ref{Table:cases}), }}}  half an our after a CCSN explosion {{{ with an explosion energy of $E_{\rm SN}=4 \times 10^{51} \erg$, }}} as we described in section \ref{subsec:3Dhydro}. The interaction of the jets with the ejecta inflates two opposite bubbles (Figs. \ref{fig:six_maps} and \ref{fig:four_maps}), similar in shapes to the bubbles that late jets, many hours after explosion, inflate in the ejecta of CCSNe \citep{AkashiSoker2021}. 

In our earlier paper the jets were active at later times and were more energetic, with energy of $E_{\rm 2j} \ge 1.6 \times 10^{49} \erg$. Here, with the induced backflow accretion, the energy in the original jets together with the jets that  the backflow accretion process launches can be as low as $E_{\rm 2j} \simeq 3 \times 10^{46} \erg$ (last column of table \ref{Table:cases}). 
We next find the inner region that the jets can influence. 

The kinetic energy of the slow ejecta with a velocity of $v < v_{\rm ej,d}$, where here we take $v_{\rm ej,d} <v_{\rm br}$, is  
\begin{eqnarray}
\begin{aligned}
& E_{\rm ej} (<v_{\rm ej,d}) = \int^{v_{\rm ej,d} t}_0 \rho_0 \left( \frac{r}{t_{\rm e} v_{\rm br}} \right)^{-1} \frac{1}{2} v^2 4 \pi r^2 dr 
\\ &
= \frac{5}{9} E_{\rm SN}  
\left( \frac{v_{\rm ej,d}} {v_{\rm br}} \right)^4  
 = 2.2 \times 10^{47}    
\left( \frac{v_{\rm ej,d}}{1000 \km \s^{-1}} \right)^4 
\\ &  \times 
\left( \frac{v_{\rm br}}{10^4 \km \s^{-1}} \right)^{-4}
\left( \frac{E_{\rm SN}}{4 \times 10^{51} \erg} \right) \erg.
\end{aligned}
\label{eq:Eej(Vejd)}
\end{eqnarray}
The mass in this inner ejecta is 
 \begin{eqnarray}
\begin{aligned}
& M_{\rm ej} (<v_{\rm ej,d}) = \int^{v_{\rm ej,d} t}_0 \rho_0 \left( \frac{r}{t_{\rm e} v_{\rm br}} \right)^{-1} 4 \pi r^2 dr 
\\ &
= \frac{7}{9} \left( \frac{v_{\rm ej,d}} {v_{\rm br}} \right)^2 M_{\rm ej} = 0.039 
\left( \frac{v_{\rm ej,d}}{1000 \km \s^{-1}} \right)^2 
\\ &  \times 
\left( \frac{v_{\rm br}}{10^4 \km \s^{-1}} \right)^{-2}
\left( \frac{M_{\rm ej}}{5 M_\odot} \right) 
M_\odot. 
\end{aligned}
\label{eq:Mej(Vejd)}
\end{eqnarray}
{{{ Note that here we take the velocity of the ejecta as it should be after the explosion, $v(r)=r/t$. Namely, here we do not include the extra numerical velocity that we introduced from numerical considerations to calculate the accreted mass due to the jets activity (section \ref{subsec:3Dhydro}). }}}

The jets interact with ejecta only along and near the polar directions, and therefore they can influence the ejecta morphology to a larger distance than that for which the ejecta energy equals the jets' energy.
If we take the region of influence to be up to {{{ where the jets have $\ga 10 \%$ average influence on the ejecta (but a larger influence along and near the polar directions) }}}, i.e., an energy of 
$E_{\rm ej} (<v_{\rm ej,d}) \simeq 10 (E_{\rm 2j} + E_{\rm 2j,next})$, then the jets might shape the inner ejecta up to velocities of $v_{\rm ej,d} \simeq 1500 [(E_{\rm 2j} + E_{\rm 2j,next})/10^{47} \erg]^{1/4} \km \s^{-1}$ for the supernova parameters that we use in this study. This velocity of the ejecta includes a mass of  $M_{\rm ej} (<v_{\rm ej,d}) \simeq 0.08 [(E_{\rm 2j} + E_{\rm 2j,next})/10^{47} \erg]^{1/2} M_\odot$. 

We conclude that the jets we study here can influence the inner ejecta even if the jets of the first post-explosion jet-launching episode are very weak, like in simulations SN1 and SN2.
The situation is similar to the jets from a NS companion at a close orbit to the collapsing core that \cite{Soker2000NScompanion} 
studied. Such jets can shape the inner ejecta to have a bipolar morphology that will in particular be imprinted on the distribution of the heavy isotopes that the explosion synthesised in the silicon layer of the core. 

Our results have implications to the analysis of the distribution of elements in supernova remnants against explosion mechanisms. There is an ongoing dispute on whether the morphological structures of clumps, rings, and filaments in some supernova remnants result only from instabilities in the explosion as the delayed neutrino mechanism predicts (e.g., \citealt{Jankaetal2017, Wongwathanarat2017, Gableretal2021, Orlandoetal2021, Sandovaletal2021}), or whether in addition to the instabilities jets also play a role as the jittering jets explosion mechanism predicts (e.g., \citealt{Bearetal2017, GrichenerSoker2017, Soker2017RAA, Akashietal2018, BearSoker2018SN1987A}). 
The most recent dispute is over the supernova remnant SNR~0540-69.3, for which \cite{Larssonetal2021} adopt the delayed neutrino mechanism and argue that instabilities alone explain the rings and clumps in this supernova remnant, while \cite{Soker2022} claims for imprints of jittering jets as well as instabilities in this supernova remnant. 
{{{ The present study does not touch directly the dispute over the explosion mechanism of CCSNe. Our study adds to the rich variety of processes that jets might play in CCSNe, and by that indirectly might be used in analysing the morphologies of CCSN remnants. For example, instabilities in the explosion process itself might form clumps and filaments in the ejecta (something which we did not include in the simulations). The jets that we simulate here can shape these clumps and filaments to a bubble-like structure. Namely, the dense shell around the large bubble that we present as the red zone in the upper left panel of Fig. \ref{fig:four_maps}, might itself be composed of filaments and clumps. A long filament near the polar zone of the dense shell might acquire a ring-like shape.  }}}

In light of our results, we predict that in some cases, probably only in a small fraction of CCSNe, the very inner part of a supernovae remnant might display a bipolar structure that results from post-explosion weak jets. The regions outside this part might display the morphology of jittering jets. More detail studies of the structure will require hydrodynamical simulations to much later times and that include more ingredients, like radiative transfer.

\section*{Acknowledgments}
{{{ We thank an anonymous referee for helpful suggestions and comments. }}}
This research was supported by the Amnon Pazy Research Foundation.



\end{document}